\documentclass[twoside,twocolumn,9pt]{article}
\usepackage{extsizes}
\usepackage[super,sort&compress,comma]{natbib} 
\usepackage[version=3]{mhchem}
\usepackage[left=1.5cm, right=1.5cm, top=1.785cm, bottom=2.0cm]{geometry}
\usepackage{balance}
\usepackage{mathptmx}
\usepackage{sectsty}
\usepackage{graphicx} 
\usepackage{lastpage}
\usepackage[format=plain,justification=justified,singlelinecheck=false,font={stretch=1.125,small,sf},labelfont=bf,labelsep=space]{caption}
\usepackage{float}
\usepackage{fancyhdr}
\usepackage{fnpos}
\usepackage[english]{babel}
\addto{\captionsenglish}{%
  
}
\usepackage{array}
\usepackage{braket}
\usepackage{droidsans}
\usepackage{charter}
\usepackage[T1]{fontenc}
\usepackage[usenames,dvipsnames]{xcolor}
\usepackage{setspace}
\usepackage[compact]{titlesec}
\usepackage{hyperref}
\usepackage{amssymb}
\usepackage{amsmath}
\usepackage{bm}
\usepackage[dvipsnames]{xcolor}
\usepackage{soul}

\usepackage{braket}

\usepackage{etoolbox}
\usepackage{tikz}

\usepackage{epstopdf}

\definecolor{massimiliano}{RGB}{0,0,255}

\definecolor{cream}{RGB}{222,217,201}

\begin{document}

\pagestyle{fancy}
\thispagestyle{plain}
\fancypagestyle{plain}{
\renewcommand{\headrulewidth}{0pt}
}


\makeFNbottom
\makeatletter
\renewcommand\LARGE{\@setfontsize\LARGE{15pt}{17}}
\renewcommand\Large{\@setfontsize\Large{12pt}{14}}
\renewcommand\large{\@setfontsize\large{10pt}{12}}
\renewcommand\footnotesize{\@setfontsize\footnotesize{7pt}{10}}
\makeatother

\renewcommand{\thefootnote}{\fnsymbol{footnote}}
\renewcommand\footnoterule{\vspace*{1pt}%
\color{cream}\hrule width 3.5in height 0.4pt \color{black}\vspace*{5pt}} 
\setcounter{secnumdepth}{5}

\makeatletter 
\renewcommand\@biblabel[1]{#1}            
\renewcommand\@makefntext[1]%
{\noindent\makebox[0pt][r]{\@thefnmark\,}#1}
\makeatother 
\renewcommand{\figurename}{\small{Fig.}~}
\sectionfont{\sffamily\Large}
\subsectionfont{\normalsize}
\subsubsectionfont{\bf}
\setstretch{1.125} 
\setlength{\skip\footins}{0.8cm}
\setlength{\footnotesep}{0.25cm}
\setlength{\jot}{10pt}
\titlespacing*{\section}{0pt}{4pt}{4pt}
\titlespacing*{\subsection}{0pt}{15pt}{1pt}

\fancyfoot{}
\fancyfoot[LO,RE]{\vspace{-7.1pt}\includegraphics[height=9pt]{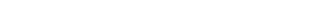}}
\fancyfoot[CO]{\vspace{-7.1pt}\hspace{13.2cm}\includegraphics{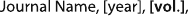}}
\fancyfoot[CE]{\vspace{-7.2pt}\hspace{-14.2cm}\includegraphics{head_foot/RF}}
\fancyfoot[RO]{\footnotesize{\sffamily{1--\pageref{LastPage} ~\textbar  \hspace{2pt}\thepage}}}
\fancyfoot[LE]{\footnotesize{\sffamily{\thepage~\textbar\hspace{3.45cm} 1--\pageref{LastPage}}}}
\fancyhead{}
\renewcommand{\headrulewidth}{0pt} 
\renewcommand{\footrulewidth}{0pt}
\setlength{\arrayrulewidth}{1pt}
\setlength{\columnsep}{6.5mm}
\setlength\bibsep{1pt}

\makeatletter 
\newlength{\figrulesep} 
\setlength{\figrulesep}{0.5\textfloatsep} 

\newcommand{\topfigrule}{\vspace*{-1pt}%
\noindent{\color{cream}\rule[-\figrulesep]{\columnwidth}{1.5pt}} }

\newcommand{\botfigrule}{\vspace*{-2pt}%
\noindent{\color{cream}\rule[\figrulesep]{\columnwidth}{1.5pt}} }

\newcommand{\dblfigrule}{\vspace*{-1pt}%
\noindent{\color{cream}\rule[-\figrulesep]{\textwidth}{1.5pt}} }

\makeatother

\twocolumn[
  \begin{@twocolumnfalse}
{\includegraphics[height=30pt]{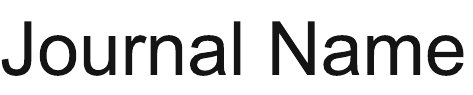}\hfill\raisebox{0pt}[0pt][0pt]{\includegraphics[height=55pt]{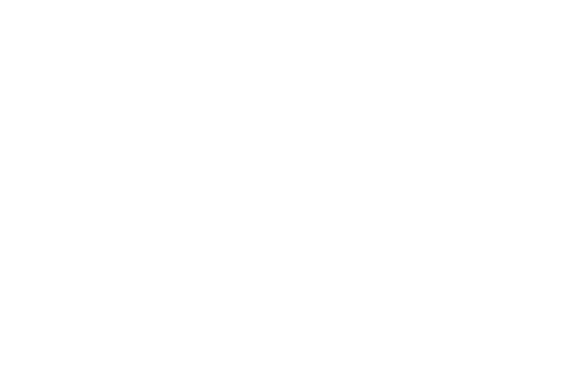}}\\[1ex]
\includegraphics[width=18.5cm]{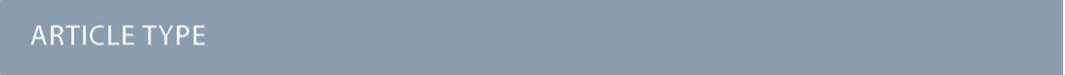}}\par
\vspace{1em}
\sffamily
\begin{tabular}{m{4.5cm} p{13.5cm} }

\includegraphics{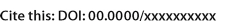} & \noindent\LARGE{
\textbf{Modelling transcriptional silencing and its coupling to 3D genome organisation~$\ddagger$}} \\
\vspace{0.3cm} & \vspace{0.3cm} \\

 & \noindent\large{M. Semeraro,\textit{$^{a}$$^{\dag}$} G. Negro,\textit{$^{b*\dag}$} D. Marenduzzo,\textit{$^{b}$}  and G. Forte\textit{$^{b**}$}} \\

\includegraphics{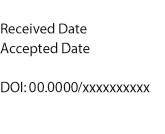} & \noindent\normalsize{Timely up- or down-regulation of gene expression is crucial for cellular differentiation and function. While gene upregulation via transcriptional activators has been extensively investigated, gene silencing remains understudied, especially by modelling. This study employs 3D simulations to study the biophysics of a chromatin fibre where active transcription factors compete with repressors for binding to transcription units along the fibre, and investigates how different silencing mechanisms affect 3D chromatin structure and transcription. 
We examine three gene silencing feedback mechanisms: positive, negative, and neutral. These mechanisms capture different silencing pathways observed or proposed in biological systems. Our findings reveal that, whilst all mechanisms lead to a silencing transition, the signatures of this transition depend on the choice of the feedback. The latter controls the morphologies of the emergent 3D transcription factor clusters, the average gene expression and its variability, or gene noise, and the network of ensuing correlations between activities of neighbouring transcription units. These results provide insights into the biophysics of gene silencing, as well as into the interplay between transcriptional regulation and 3D genome organisation. 
} \\

\end{tabular}

 \end{@twocolumnfalse} \vspace{0.6cm}

  ]

\renewcommand*\rmdefault{bch}\normalfont\upshape
\rmfamily
\section*{}
\vspace{-1cm}


\footnotetext{\textit{$^{a}$~Dipartimento Interateneo di Fisica, Università degli Studi di Bari and INFN, Sezione di Bari, via Amendola 173, Bari, I-70126, Italy}}
\footnotetext{\textit{$^{b}$~SUPA School of Physics and Astronomy, University of Edinburgh, Peter Guthrie Tait Road, Edinburgh EH9 3FD, UK}}

\footnotetext{\dag~These authors contributed equally to this work.}
\footnotetext{$\ast$~{giuseppe.negro@ed.ac.uk}}
\footnotetext{$\ast$$\ast$~{gforte@ed.ac.uk}}
\footnotetext{$\ddagger$~ Electronic supplementary information (ESI) available: 
Details on the polymer model we adopt, simulation and sampling as well as  supplementary figures are included in a pdf file. See DOI: xxx}


\section{Introduction}

During interphase, chromatin, the fibre composed of DNA and histone proteins, is transcribed into mRNA, which in turn provides the template to assemble proteins~\cite{Alberts2017}. Although the whole genetic material is present in every cell, only a subset of genes is transcribed in each cell type, leading to what is known as cellular differentiation~\cite{Slack2021}. 
Experiments investigating how transcription is orchestrated keep exposing exciting challenges to address and new questions to answer~\cite{Dupont2022, Grosveld2021}. A leitmotif is the  apparent link between the regulation of gene expression and 3D chromatin organisation. The advance of experimental techniques has allowed to find indirect evidence for this connection~\cite{Pombo2015, Ibrahim2020,du2024}.\\
Hi-C experiments, detecting contacts between distant genomic loci~\cite{Vsimkova2024}, have revealed the presence of topologically associated domains - usually sub-megabase chromatin regions enriched in self interactions - that often  contain co-regulated genes~\cite{Ibrahim2020,Long2022}. At a larger scale, chromatin is organised into A and B compartments, which closely correspond to euchromatin and heterochromatin regions first observed by optical microscopy in the 1930s~\cite{Liu2020}. Heterochromatin, typically located at the  nuclear periphery, appears as  highly condensed chromatin, whereas euchromatin, generally occupying more central regions of the nucleus, is less compact. It is now established that euchromatin, being more accessible to transcription factors and RNA polymerases, predominantly contains actively transcribed genes, while genes within heterochromatin domains are largely associated with transcriptional repression~\cite{Hildebrand2020}. \\
Gene transcription - initiated and maintained by interactions between RNA polymerases, active transcription factors, and chromatin elements such as promoters and enhancers\cite{Panigrahi2021, Sahu2022}- has been extensively investigated both experimentally~\cite{Marguerat2010,Schwalb2016,Cheng2023} and through theoretical modeling~\cite{brackley2021,Semeraro2025,iscienceNegro}. In contrast, comparatively less is known about gene silencing - the mechanisms by which gene transcription is turned off~\cite{Pang2020}. \\
In bacteria, gene silencing is typically mediated by transcription repressor proteins that inhibit gene expression by binding to the operator, a specific genomic region located near the promoter. This binding prevents RNA polymerase from accessing the promoter, thereby blocking transcription.~\cite{Thiel2004}. A well-known example is provided by the lac operon repressor, which, in the absence of lactose, hinders the loading of RNA polymerases on lac operon promoter~\cite{Oehler1990}.\\
In eukaryotes, transcription silencing is  more convoluted and it encompasses a  wide spectrum of mechanisms. First, eukaryotic gene silencing can be achieved through regulatory elements known as silencers, which serve as binding sites for repressor proteins~\cite{Alberts2017}. When a repressor binds to a proximal silencer -- located near a gene promoter -- it reduces the likelihood of RNA polymerase loading onto the promoter, analogous to transcription repression in  bacteria~\cite{Cooper2022}. Alternatively, an eukaryote repressor may simultaneously bind  to a distal silencer and to a promoter, resulting in the promoter becoming inaccessible to RNA polymerase and other activator proteins~\cite{Zhang2022}.
Second, eukaryotes can repress transcription by means of local alterations in chromatin structure~\cite{Beisel2011,Cooper2022, Lodish2004}. ChIP-seq techniques have shown that specific DNA and histone modifications, known as epigenetic marks, are closely associated with distinct chromatin states. For example, trimethylation of histone H3 on lysine 27 (H3K27me3) is linked to heterochromatin formation, whereas histone H3 lysine 27 acetylation (H3K27ac) tags more accessible and transcriptionally active chromatin regions. Gene silencing can thus be mediated  by the deposition and removal of these epigenetic marks, often carried out by  corepressor complexes - protein assemblies recruited to chromatin by intermediate proteins - that function as epigenetic writers~\cite{Perissi2010}.  For instance, histone acetyl groups are  removed by protein complexes carrying the enzymes HDAC1  and HDAC2~\cite{Sengupta2004}, such as the corepressor coREST, which is recruited to chromatin via interaction with transcription factors like REST~\cite{Andres1999}. \\ 
Although the importance of eukaryotic gene silencing, alongside activation, is now emerging through new experimental findings~\cite{Zhang2021,Capriotti2020}, a mechanistic understanding of the balance between  positive and negative  transcription regulation remains elusive~\cite{Blackledge2021, Wong2020}. Little is known about the quantitative impact of    different repression mechanisms on gene activity and how they may alter the 3D structure of active chromatin, for example, by disrupting transcription factories - clusters of active RNA polymerases and associated transcription factors~\cite{Cooper2022,Papantonis2013}. 

To gain new insights into the biophysical mechanisms of gene silencing, here we develop a polymer model based on 3D coarse-grained molecular dynamics simulations. In our framework, chromatin is represented as a polymer interacting with diffusing active and repressive transcription factors, enabling us to investigate how distinct silencing mechanisms affect the composition of transcription factor clusters and, consequently, gene expression. 
Although our approach is based on a simplified toy model that includes only essential components, it is important to note that models similar in spirit to ours - based on sequence and average epigenetic marks - have been successfully used in the past 
to explain the formation of transcription factories, and to predict chromatin loops and contact maps consistent with experiments~\cite{Barbieri2012,Jost2014,DiPierro2016,Zhan2017,Bianco2018,brackley2016, Merlotti2020, brackley2021, semeraro2023}.
Compared to these earlier models, ours includes dynamic deposition and removal of repressive epigenetic marks. 
This allows us to explore, for the first time, the interplay between active transcription factors and repressors in a unified, dynamic setting. Our findings offer predictions that may inform  future experiments and guide the development of more sophisticated models investigating the spatio-temporal dynamics of transcription. 

\section{Model}

\begin{figure}[t!]
\centering
\includegraphics[width=0.95\columnwidth]{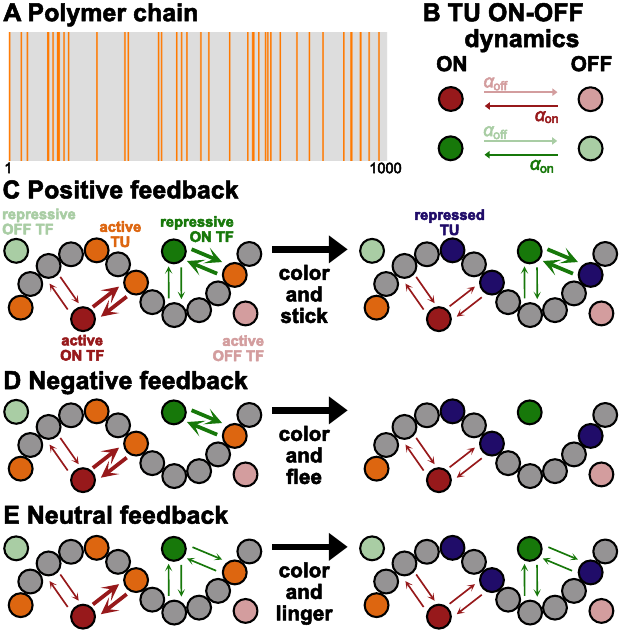}
\caption{\footnotesize{(A) Initial polymer bead sequence. Orange and gray lines denote the location of active TUs and unmarked chromatin sites respectively. Bead ID number ranges between $1$ and $1000$. (B) Sketch of the {\it ON-OFF} dynamics of active (red) and repressive (green) TFs occurring with switching rates $\alpha_{on}=\alpha_{off}$. TFs in the \textit{OFF} state only interact sterically with the chromatin filament. Interactions between chromatin and \textit{ON} TFs depend on the silencing feedback mechanism. (C) to (E) Schematic representation of the three silencing models (positive, negative and neutral feedback). The chromatin filament is represented as a polymer formed by a sequence of connected beads. Active and repressed TUs are depicted as orange and blue beads respectively, while unmarked chromatin is represented by gray beads. {\it ON} and {\it OFF} active (repressive) TFs are instead depicted as dark and light red (green) beads respectively. Thicker and thinner arrows denote strong and weak attractive interactions. In the {\it positive feedback} model (panel (C)) repressive {\it ON} TFs are strongly attracted to both active and repressed TUs, and weakly to unmarked chromatin beads. In the {\it negative feedback} model (panel (D)) repressive \textit{ON} TFs are strongly attracted only to active TUs, while in {\it neutral feedback} (panel (E)) they are weakly attracted to the whole chromatin filament. In all three feedback mechanisms active TFs are strongly attracted to active TUs and experience a weak attraction to repressed TUs and unmarked chromatin.}}
\label{fig:fig1}
\end{figure}

Chromatin is depicted as a coarse-grained polymer, whose beads are assumed to contain $1-3~kbp$ each, corresponding to a bead diameter $\sigma \sim 20-30~nm$~\cite{Chiang2022_book}. Consecutive beads are connected via spring potentials and an additional Kratky-Porod-like potential applied to triplets of consecutive beads is used to endow the filament with a persistence length. The chromatin polymer is initially composed of two types of beads representing non-specific sites and active transcription units, TUs (grey and orange beads in Fig.~\ref{fig:fig1}(C)). The latter can be thought of as gene promoters or enhancers. For the sake of comparison, we consider the same $1000$ beads long polymer chain as in Ref.~\cite{brackley2021, semeraro2023} with $39$ TUs placed randomly as depicted in Fig.~\ref{fig:fig1}(A). Transcription factors (TFs), which are simulated as additional spheres diffusing in the system and interacting with chromatin, are divided into active and repressive (red and green beads in Fig.~\ref{fig:fig1}). The latter have the ability of silencing a TU they are bound to with a given probability $p_s$, generating a repressed TU (blue beads in Fig.~\ref{fig:fig1}), which can revert back to an active TU after a time $\tau_{R}$. As depicted in Fig.~\ref{fig:fig1}(B), both active and repressive TFs switch back and forth between an \textit{ON} and \textit{OFF} state at rates $\alpha_{on}=\alpha_{off}$ with a probability $p_{switch}$. Similarly to Ref.~\cite{brackley2021, semeraro2023}, we consider $40$ active TFs and as many repressive TFs, with $20$ of them initially in the initial \textit{ON} state and the remaining in the \textit{OFF} state. When in the \textit{OFF} state, active and repressive TFs experience a purely steric interaction with the whole chromatin filament. Active \textit{ON} TFs are strongly attracted to active TUs and weakly to all other chromatin beads (non-specific sites and repressed TUs). The interaction between repressive \textit{ON} TFs and chromatin depends instead on the transcriptional feedback we simulate. We consider three different schemes, or transcriptional feedbacks, each modelling a different silencing mechanism. 

In the {\it positive feedback} model (Fig.~\ref{fig:fig1}(C)), repressive \textit{ON} TFs experience a strong affinity for both active and repressed TUs, while displaying a weaker attraction to non-specific binding sites. This represents a  \textit{color and stick} scenario, wherein repressive TFs bind to active TUs, repress them and subsequently maintain a high affinity for repressed TUs. As a result, the latter are less likely to regain accessibility and to become attractive to active TFs. This mechanism is analogous to the repression observed in bacteria, where a repressor physically obstructs the binding of RNA polymerase to promoter regions. \\
The {\it negative feedback} model (Fig.~\ref{fig:fig1}(D)) involves repressive \textit{ON} TFs sterically interacting with the majority of chromatin beads, with the exception of  active TUs, to which they bind with high affinity. This corresponds to  a \textit{color  and flee} scenario, as repressive TFs typically detach from chromatin after silencing active TUs (and hence turning them into inactive TUs). This feedback simulates the action of  proteins that modulate chromatin accessibility and are commonly associated with repressors. An example is the previously mentioned corepressor coREST, which carries the enzymes HDAC1 and HDCA2, leading to the removal of acetylations marks and, consequently, chromatin compaction and transcription repression.\\
Finally, in the {\it neutral feedback} model(Fig.~\ref{fig:fig1}(E)), repressive \textit{ON} TFs  are weakly attracted to the entire chromatin filament, regardless of bead type. This represents a \textit{color and linger} scenario, in which  repressive TFs keep hanging around the chromatin chain, sliding along it, even after silencing active TUs. This mechanism simulates an alternative mode of interaction for corepressors involved in deacetylation, which may display low rather than no affinity for unmarked chromatin and repressed TUs.

We will now investigate the effects of these three mechanisms on 3D chromatin  organisation and transcriptional activity. The simulation box we consider is large enough to ensure that the system is dilute. More details about the model, its implementation and the sampling of observables are reported in Ref.~\cite{SM}.


\section{Results}


\subsection{Size and composition of clusters of transcription factors}

\label{sec:clustering}

\begin{figure*}[t!]
\includegraphics[width=1.95\columnwidth]{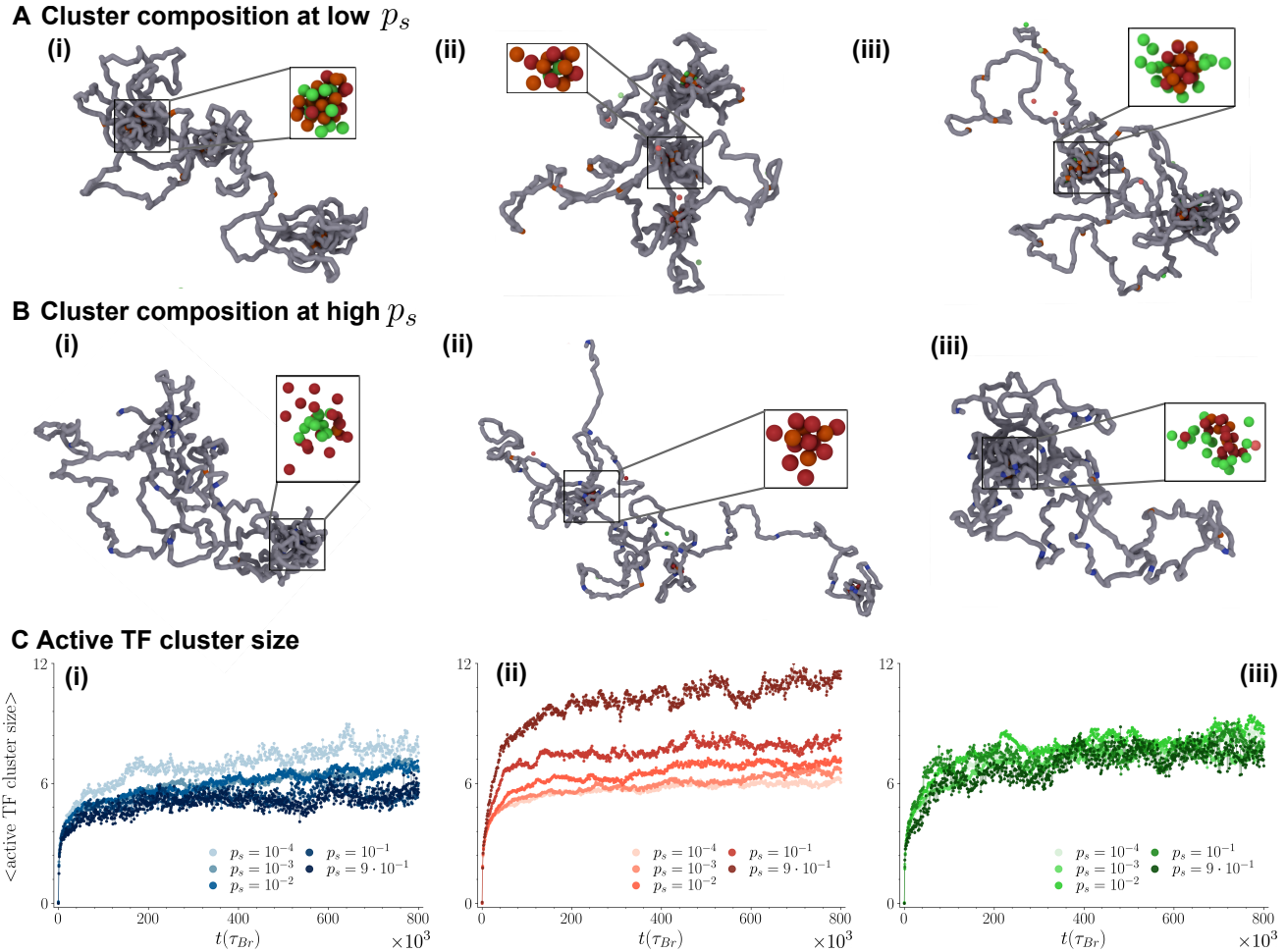}
\caption{\footnotesize{(A) Simulation snapshots showing the typical $3D$ arrangement of the chromatin filament and TF cluster structure (insets) at low silencing probability ($p_s=10^{-4}$). Panels (i), (ii) and (iii) refer to the three silencing mechanisms assumed in the models (positive, negative and neutral feedback respectively). (B) Typical simulation snapshots for the positive, negative and neutral feedback ((i), (ii), (iii)) at high silencing probability ($p_s = 0.05$ in (i) and (iii), $p_s = 0.1$ in (ii)). (C) Average size of clusters formed by active TFs over time for the the positive (panel (i)), negative (panel (ii)) and neutral (panel (iii)) feedback models. The darker the curve, the higher the silencing probability $p_s$ it represents. }}

\label{fig:fig_clusters}
\end{figure*}

Since simulated transcription factors represent complexes of proteins and RNA Polymerases, we model them as multivalent spheres capable of simultaneously binding to multiple chromatin loci, thereby forming bridges. Notably, their multivalent nature, combined with their ability to engage in both weak and strong attractive interactions with the chromatin polymer, drives the spontaneous formation of TF clusters through a mechanism known as bridging-induced phase separation (BIPS)~\cite{Brackley2013, Brackley2020}. In a simplified scenario,  when a multivalent TF, attracted to the entire chromatin filament, binds to a chromatin bead, it induces a local increase in chromatin density by attracting additional nearby chromatin beads. This process, in turn, facilitates the binding of additional TFs in the vicinity of the initial TF, where the density of chromatin binding sites has increased, leading to the formation  of a TF cluster. \\
\begin{figure*}[t!]
\includegraphics[width=1.95\columnwidth]{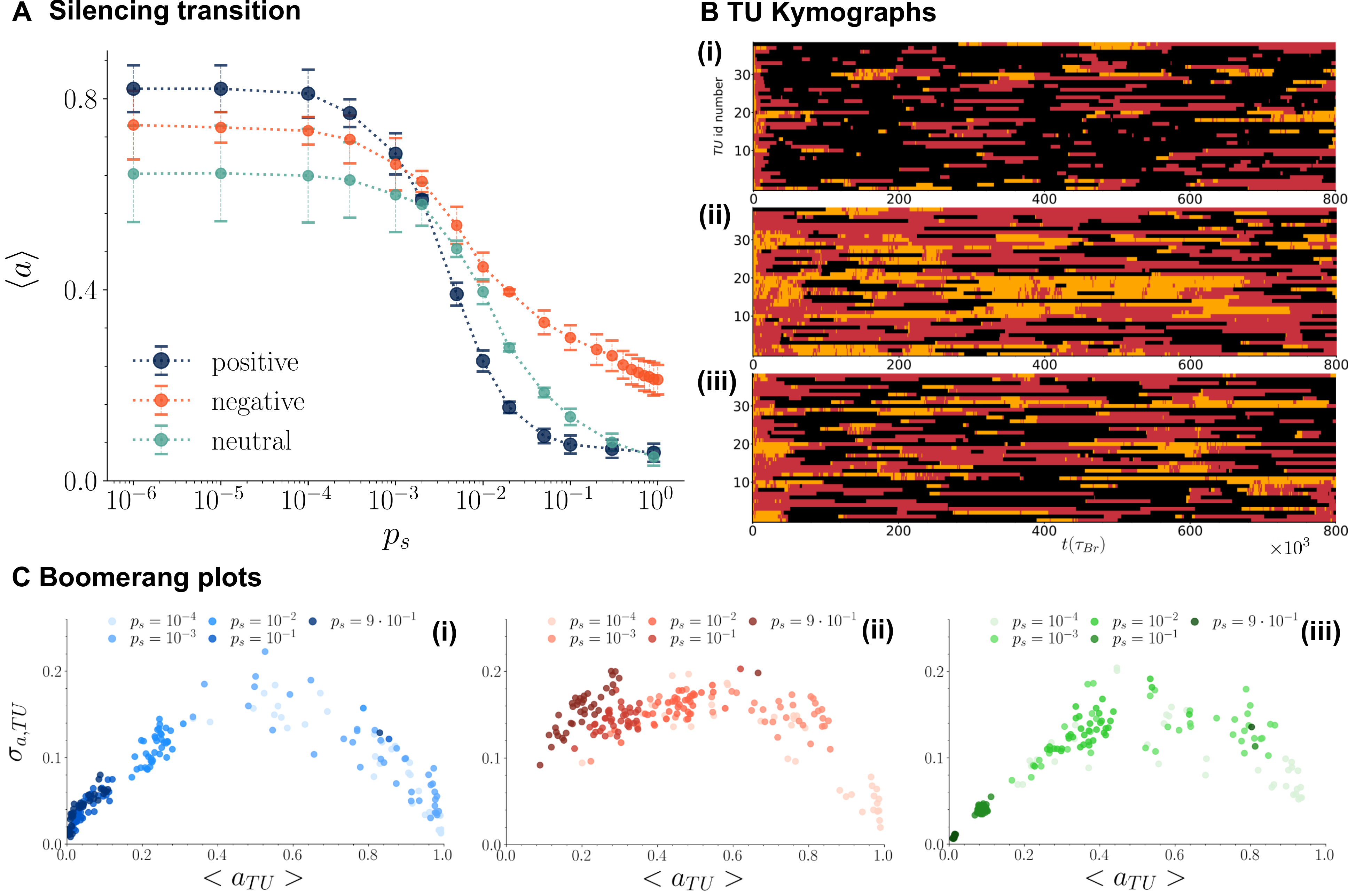}
\caption{\footnotesize{(A) Average transcriptional activity of active TUs $<a>$ versus silencing probability $p_s$ for the positive (blue curve), negative (red curve) and neutral (green curve) feedback mechanisms. For each value of $p_s$ and for each feedback case, $<a>$ values are obtained by first computing the average of the activity of each TU over $100$ independent simulations, $<a_{TU}>$, and then averaging the values of $<a_{TU}>$ over the different TUs. All three silencing mechanisms show a drop in  $<a>$ above a critical $p_s$, lying between $10^{-3}$ and $10^{-2}$. We refer to this phenomenon as the {\it silencing transition}.  (B) Kymographs of a single representative run with $p_s=10^{-2}$ for the positive (panel (i)), negative (panel (ii)) and neutral (panel (iii)) feedback case. In the y-axis the ID number of each of the $39$ TUs in the chromatin chain is reported (the ID number ranges between $1$ and $1000$ as the chromatin polymer is composed of $1000$ beads). Time is shown in the x-axis. Black, yellow and red pixels denote the transcriptional state of a specific TU at a certain time point. Black pixels refer to \textit{OFF} TUs, yellow pixels refer to non-transcribing \textit{ON} TUs (i.e. \textit{ON} TUs without any active TF close by), while red pixels correspond to actively transcribing \textit{ON} TUs. (C) ``Boomerang plots''~\cite{chiang2024} showing the variance of the activity $\sigma_{a,TU}$ versus the  average activity $<a_{TU}>$ for each TU of the chromatin filament. Darker spots represent larger $p_s$ values. Averages and variances are computed over $100$ independent simulations. Panels (i), (ii) and (iii) refer to the three silencing mechanisms (positive, negative and neutral feedback respectively).}}
\label{fig:activity}
\end{figure*}
The formation and composition of TF clusters appear to be significantly affected by both the silencing mechanism assumed in the model and the silencing probability $p_s$. Fig.~\ref{fig:fig_clusters}(A-B) illustrates representative cluster structures observed under low and high $p_s$ conditions across the  three transcriptional feedback mechanisms. At low  $p_s$ values, only a few TUs are repressed, resulting in cluster composition being primarily determined by interactions between TFs and active TUs. In the \textit{color and stick} model, active and repressive TFs experience the same attraction with active TUs and non-specific chromatin beads, leading them to form mix clusters 
(Fig.~\ref{fig:fig_clusters}(A,i)). In the other two mechanisms, TFs still aggregate around active TUs, but active and repressive TFs no longer mix. In the \textit{color and flee} scenario, active TFs bind both to active TUs (strongly) and unmarked chromatin sites (weakly), while repressive TFs only bind to active TUs: as a result, repressive TFs are driven to cluster cores, which are enriched in TUs, while active TFs localize at outer shells of clusters, where they can bind both TUs and the nearby unmarked chromatin (see Fig.~\ref{fig:fig_clusters}(A,ii)). Conversely, in the \textit{color and linger} mechanism,  repressive TFs  exhibit only  weak attraction to active TUs, driving them to the outer layer of the TF clusters (Fig.~\ref{fig:fig_clusters}(A,iii)). \\
A sharp increase in the silencing probability $p_s$ results in a significantly larger number of repressed TUs. 
In the \textit{color and stick}  mechanism, the composition of  clusters undergoes a notable shift: since active and repressive TFs experience weak and strong  interactions with repressed TUs respectively, active TFs now relocate to the outer shell of  TF clusters, while repressive TFs form the cluster core (Fig.~\ref{fig:fig_clusters}(B,i)). Furthermore, active TFs are  less likely to remain stably bound to  a cluster, leading to a decrease in the average size  of clusters formed by active TFs  as $p_s$ increases (Fig.~\ref{fig:fig_clusters}(C,i)). In the \textit{color and flee} scenario, clusters are primarily composed of active TFs, as repressive TFs  dissociate from the chromatin filament after it is recolored ( Fig.~\ref{fig:fig_clusters}(B,ii). However,  the reduction in active TUs sites leads to a decrease in the number of active TF clusters and a concomitant increase in their average  size ( Fig.~\ref{fig:fig_clusters} (C,ii)). Finally, the \textit{color and linger} mechanism combined with  a high number of repressed TUs (i.e., high $p_s$) involves similar attractive interactions between  the chromatin filament and both  active and repressive TFs, which then form mixed clusters. 
Here, the weak attraction between (active and repressive) TFs and the repressed TUs, combined with the strong interaction between active TFs and remaining active TUs is sufficient to maintain the active clusters stable (Fig.~\ref{fig:fig_clusters}(C,iii)). \\

\subsection{The silencing transition}
As previously mentioned, the spatial arrangement of chromatin in 3D is closely linked to the transcriptional activity. Therefore, it is not surprising that the variety of structures and compositions of TF clusters observed in our simulations correlates with differences in the transcription levels of active transcription units. \\
From a simulation perspective, the transcription activity of an active TU can be predicted by measuring the fraction of time the TU is bound by an active TF~\cite{brackley2021}. In Fig.~\ref{fig:activity}(A), the transcription activity $<a>$, averaged over all active TUs in the chromatin fibre, is plotted as a function of $p_s$ for the three transcriptional feedback cases. The three curves show a sharp decrease in $<a>$ for critical values of $p_s$ between $10^{-3}$ and $10^{-2}$, which  we refer to  as the \textit{silencing transition}. At low silencing probabilities, the positive feedback model (\textit{color and stick}) exhibits the highest transcriptional activity, reflecting the formation of large clusters composed of both active and repressive TFs, which incorporate multiple active TUs and enhance their transcription activity (see Fig.S2 in ESI$\ddagger$).
On the other hand, the neutral feedback model (\textit{color and linger}) shows the lowest activity at small $p_s$, as the repressive TF outer shell in the clusters prevents nearby active TUs from joining the cluster and contributing to the overall activity (see Fig.~\ref{fig:fig_clusters}(A,iii)). Interestingly,  after the silencing transition -- at higher $p_s$ values -- the negative feedback model (\textit{color and flee}) displays the highest activity, reflecting the formation of larger clusters of active TFs (Fig.~\ref{fig:fig_clusters}(C,ii)). In contrast, the positive feedback model  undergoes  a sharp silencing transition, as the large clusters of both active and repressive TFs observed at small $p_s$ dissolve, giving rise to  clusters composed mainly of repressive TFs  (see Fig.~\ref{fig:fig_clusters}(B,i)). In all cases, therefore, the average transcriptional activity  can be understood by inspecting the composition and size of TF clusters, providing another example of a link between 3D chromatin structure and transcription.\\
Besides having an effect on the {\it average} transcription activity among all TUs, the three different transcriptional feedback mechanisms yield notable differences on {\it single} TU transcriptional activity (see Fig.S3 in ESI$\ddagger$) and on transcription {\it noise} (or variability). Kymographs in Fig.~\ref{fig:activity}(B) show the activity state of each TU after the silencing transition ($p_s = 10^{-2}$). Black, yellow and red pixels respectively correspond to repressed TUs, active non-transcribing TUs (i.e., without an active TF close by) and active transcribing TUs. The latter are clearly more frequent in the negative feedback, where larger clusters of active TFs form enhancing the activity. More kymographs at different $p_s$ are reported in Fig.S4$\ddagger$. A quantification of the transcriptional variability of single TUs is instead shown in Fig.~\ref{fig:activity}(C). The variability is normally referred to as transcriptional noise~\cite{chiang2024}, and provides an indication about how the activity of a single TU changes from cell to cell, because, for instance, of molecular events such as binding-unbinding  of TFs, 
or due to the variability of TF concentration inside a cell~\cite{semeraro2023, Chiang2022_arxiv}. From a simulation point of view, the noise of a given TU, $\sigma_{TU}$, is computed as the standard deviation of its activity averaged over independent simulations. 
As  each  simulation can be thought as representing a single cell, the transcriptional noise obtained from a set of simulations can be related to the variation of transcriptional activity that would be observed in a cell population. By plotting $\sigma_{TU}$ versus $<a_{TU}>$ for each TU and  at different values of $p_s$, we obtained what we call a \textit{boomerang plot}~\cite{chiang2024} (Fig.~\ref{fig:activity}(C)). For the three mechanisms, the activity is larger for small values of $p_s$ -- as  TUs are more likely to be active and transcribing -- and it decreases with the increase of the silencing probability. Interestingly, in the \textit{color and stick} and \textit{color and linger} models, the noise $\sigma_{TU}$ reaches its maximum close to the silencing transition -- this is similar to what is expected to happen near a phase transition, where fluctuations peak. 
In the \textit{color and flee} feedback, instead, $\sigma_{TU}$ remains high even after the transition ( for $p_s \geq 10^{-3}$), as the lack of attraction between repressive TFs and repressed TUs does not favour the formation of repressive TF clusters around repressed TUs, which can then easily revert back to an active state, increasing the transcription variability (Fig.~\ref{fig:activity}(C,ii)). \\

\subsection{Silencing-induced patterns of transcriptional correlations}

\begin{figure}[t!]
\includegraphics[width=0.99\columnwidth]{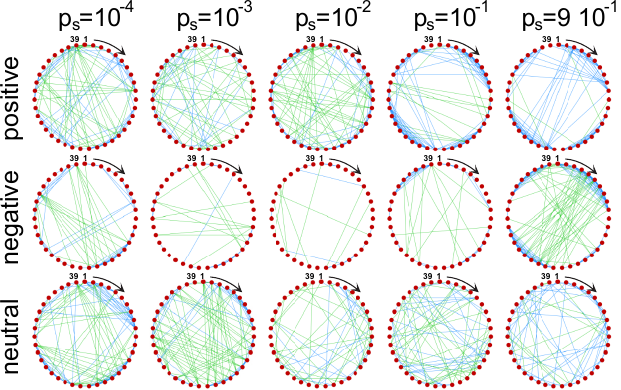}
\caption{\footnotesize{TU correlation networks for five values of $p_s$ (columns) in the positive, negative and neutral feedback mechanism (top, middle and bottom row respectively). The complete networks consist of $n=39$ individual TUs. TUs from first to last are shown as peripheral nodes and run clockwise from top. Blue and green edges respectively denote positive and negative correlations above a threshold of $0.25$, corresponding to a $p$-value $\sim2~10^{-2}$. 
We show in Ref.~\cite{SM} that most of the interactions (edges) are statistically significant.
}}
\label{fig:correlations}
\end{figure}

Fig.~\ref{fig:activity} shows that the silencing transition is accompanied by the changes in structure and composition of TF clusters observed while varying $p_s$ in Fig.~\ref{fig:fig_clusters}. But this is not the whole story: as we shall see, 
the silencing transition also drives a change in the spatiotemporal patterns of transcription of different TUs.

This phenomenon can be see from Fig.~\ref{fig:correlations}, where red dots represent the different TUs composing the chromatin chain, while blue and green lines connecting two TUs indicate a positive and negative correlation between their transcription activity, respectively (the Pearson correlation matrices used to obtain Fig.4 are shown in Fig.S5 $\ddagger$). The emerging correlation networks are visibly affected by the value of $p_s$, i.e. by the point along the silencing transition in Fig.~\ref{fig:activity}(A). For the three feedback scenarios, a small $p_s$ results in the formation of mostly positive short range correlations, due to the formation of clusters of active TFs, and negative long range correlations, unveiling the competition between active TUs for a finite number of active TFs. This is similar to the multicolour model studied in Ref.~\cite{Semeraro2025}, where no epigenetic feedback was considered. 

However, the trend of correlation networks is different in the three feedback scenarios as the system goes through the silencing transition. First, in the positive feedback (color and stick) model, positive short range correlations fade away while approaching the transition, and reappear again at high values of $p_s$. At $p_s=9 \cdot 10^{-1}$,  positive correlations are observed at both short and long range, indicating that these are generated by the formation of stable clusters of repressive TFs which switch off the transcription activity of the TUs they bind to. In the positive feedback case, then, the positive correlations we observe are related to clusters of active TUs before the silencing transition (small $p_s$) and to clusters of repressed TUs after it (high $p_s$). Second, in the negative feedback (color and flee) model, positive short range correlations also disappear while going through the transition and form again at high $p_s$, but in this case they are always led (at any value of $p_s$) by the formation of clusters of active TFs which results in negative long range correlations. Finally, in the neutral feedback (color and linger) model, after the transition TUs are mainly positively correlated, both at short and long range. 

\section{Conclusions}
\label{sec:concl}

In summary, we have presented a new model coupling 3D chromatin structure to transcriptional silencing, to investigate the potential mechanisms of eukaryotic gene repression, and the signatures they leave on quantities which can in principle be measured experimentally, such as average gene expression, transcription factory composition, and emergent transcriptional correlation networks. Our model includes two species of transcription factors, one promoting gene transcription (active TFs) and the other promoting gene silencing (repressive TFs). Through 3D coarse grained molecular dynamics simulations, we have shown that different silencing mechanisms 
influence the nature of the silencing transition, as well as the size and composition of the emergent clusters of transcription factors. First, a 
positive feedback (color and stick) silencing mechanism, through which repressors bind and modify transcription units, results in a silencing transition, from mixed clusters of active TFs and repressors to silenced clusters of tightly-bound repressors which prevent gene activation. Second, a negative feedback (color and flee) silencing mechanism, through which repressors bind strongly active transcription units but do not bind at all inactive ones, gives a silencing transition accompanied by the formation of significantly fewer, albeit larger, active TF clusters. Finally, a neutral feedback (color and linger) silencing mechanism, through which repressors inactivate transcription units without binding them strongly, leads to a shallower silencing transition between active and mixed clusters of similar size. 


We also find that the variability in the transcriptional activity of single genes, also known as transcriptional noise, is maximum near the silencing transition for the positive and neutral feedbac: this is because TUs are more likely to switch back and forth between an active and silenced state close to the critical point (Fig.~\ref{fig:activity}(C)). Instead, in the negative feedback model, which may be appropriate to the case of eukaryotic co-repressors containing writers or erasers of epigenetic marks, the lack of non-specific interactions between chromatin and silencing protein complexes results in a large noise even after the silencing transition. The control of transcriptional noise is important in differentiation, as controlling the variability of expression
may enable cells to adapt gene expression during development or in response to environmental signals. 

Finally, the three silencing feedback mechanisms investigated in this work also shape the  network of correlations between the activity of different TUs: following the silencing transition, long-range correlations become positive in the positive feedback model, while they keep being negative in the negative feedback model, where the absence of non-specific interactions is once again key 
(Fig.\ref{fig:correlations}). \\
In the future, it would be interesting to include the action of repressors into more complex and realistic chromatin models, such as the HiP-HoP model~\cite{Buckle2018,chiang2024cellgenomics}, which incorporate more chromatin compaction levels depending on epigenetic marks and the action of cohesin-driven loop extrusion~\cite{Chiang2022_book,Semeraro2025}. By investigating the correlation of simulated gene activity with experimental data inferred from GRO-seq or RNA-seq methods~\cite{Danko2015, Marguerat2010} it should be possible to gain a deeper understanding of the importance played by silencers in gene expression and chromatin architecture. Additionally, experiments involving  live-cell imaging providing information about the binding sites of silencer complexes could help test our findings which relate gene activity to TF cluster composition.

\section*{Acknowledgements}
Numerical calculations have been made possible through
a Cineca-INFN agreement, providing access to HPC resources at
Cineca. M.S. acknowledges support from INFN/FIELDTURB project and from MUR projects PRIN 2020/PFCXPE, PRIN 2022 PNRR/P20222B5P9 and Quantum Sensing and Modelling for One-Health (QuaSiModO). D.M and G.F. acknowledge support from the Wellcome Trust (223097/Z/21/Z). G.F. acknowledges support from the Leverhulme Trust (Early Career Fellowship ECF-2024-221). For the purpose of open access, the author has applied a Creative Commons Attribution (CC BY) licence to any Author Accepted Manuscript version arising from this submission.

\section*{Conflicts of interest}

    There are no conflicts to declare. 

\bibliographystyle{unsrt}
\bibliography{biblio_sm}

\end{document}